\documentstyle[12pt,epsfig]{article}
\begin{document}

\begin{center}
{\LARGE Scheme Independence of $g_1^p (x, Q^2)$}
\end{center}
\vspace{0.5cm}
\begin{center}
\begin{large} 
F. M. Steffens\footnote{fsteffen@if.usp.br},
\\
\end{large}
Instituto de F\'{\i}sica, USP, C. P. 66 318, 05315-970,SP, Brasil 
\\
\end{center}

\begin{abstract}
We work with two general factorization schemes in order
to explore the consequences of imposing scheme independence 
on $g_1^p (x, Q^2)$. We see that although the light quark sector
is indifferent to the choice of a particular scheme,  
the extension of the calculations to the heavy quark
sector indicates that a scheme like the $\overline{MS}$ is
preferable.
\end{abstract}

The problem raised by the results of the EMC spin experiment 
\cite{emc88} was deeply influential on a substantial part of the 1990's 
research in both theoretical and experimental hadron and particle physics. 
Their data implied that the quark singlet axial charge measured in a
proton target, $g_a^0$,  was compatible with zero, while quark model 
calculations predicted $g_a^0$ to lie in the range of 0.6 - 0.7.
After a series of experiments made at CERN, SLAC and HERA over the past
10 years, it is accepted today that $g_a^0 \approx 0.3$, which is
still far from those early theoretical expectations.

In 1974, Ellis and Jaffe proposed a sum rule \cite{ellis74} for the 
integral in $x$ of $g_1^p (x)$, where they assumed that
the sea quarks in the proton are not polarized. This implies that
$\Delta s$, the helicity carried by the strange quarks, is zero. 
Experimentally, $\int_0^1 g_1^p (x, Q^2=10 \; GeV^2) dx = 0.120 \pm 0.005 
\pm 0.006 \pm 0.014$ \cite{smc98}, while the Ellis and Jaffe sum rule 
gives $\int_0^1 g_1^p (x, Q^2=10 \; GeV^2) dx = 0.176 \pm 0.006$.

In the the parton model, the first moment of $g_1^p (x)$ is given
by $\int_0^1 g_1^p (x) dx = \frac{1}{12}g_a +  \frac{1}{36}g_a^8  
+ \frac{1}{9}g_a^0$,
with $g_a = \Delta u - \Delta d$ the isotriplet axial charge,
$g_a^8 = \Delta u + \Delta d - 2 \Delta s$ the octet axial charge, and
$g_a^0 = \Delta u + \Delta d + \Delta s$. 
The parton model structure functions are, actually, QCD structure functions
with ${\cal O}(\alpha_s^0)$ corrections. 
Beyond the parton model, the identification of the singlet axial charges with 
the sum of the quark helicities ceases to be true, because of the 
clash between a gauge invariant and a chiral symmetric renormalization
procedure of the axial charge \cite{bardeen,others}.

The failure of the Ellis and Jaffe sum rule to agree with the
experiments is translated into the non-equivalence between the 
quark singlet and octet axial charges.  
From the start there has been a large controversy on the mechanism
responsible for  $g_a^0 \neq  g_a^8$. In the context of the parton 
model, $\Delta s \neq 0$ settles the question. However, as proposed
by Altarelli and Ross \cite{altarelli88}, and Carlitz, Collins and
Mueller \cite{carlitz88}, it is still possible to have $\Delta s = 0$
if one takes into account the axial anomaly \cite{adler} which appears
in the QCD calculations of $g_1^p (x, Q^2)$ at ${\cal O}(\alpha_s)$.
Later, it became clear that these two different scenarios,
$\Delta s \neq 0$ or an anomaly contribution, are simply related by a 
change of scheme defining the partons distributions and the coefficient 
functions. This will be, indeed, the main contribution of this work.
We will argue that although the appearence of gluons 
in the first moment of $g_1^p (x, Q^2)$, in the light quark sector,
is a matter of scheme preference, the introduction of heavy quarks suggests
that a scheme where the gluons do not contribute, like the $\overline{MS}$ 
scheme, is preferable. 

A part of what is discussed in this work have already been adressed in the
literature. Specifically, the importance in isolating the hard part
of the photon-gluon cross section \cite{bodwin90,lech,me,cheng96}.
However, some missconceptions still persist, mainly those connected 
with the heavy quark contribution to $g_1^p(x,Q^2)$, which is one
of the motivations for the explicit discussions made here on the 
ways to calculate a polarized 
gluon coefficient function which is free of ambiguity in the 
infrared region.

The choice of a factorization scheme is a reflection of the choice of a 
regulator and
of a subtraction for the soft and collinear divergences appearing in the
calculation of the partonic cross sections. In the specific case of the axial
anomaly contribution to $g_1^p$, much have been discussed about the ambiguity 
in the  choice of a quark or of a gluon mass to regulate these 
divergences \cite{altarelli88,carlitz88,bodwin90,lech}. 
In a satisfactory calculation, the hard part
of the partonic cross sections should not present any ambiguity. 
The infrared singularities are present in the full virtual photon-gluon
cross section \cite{bass90,vogelsang}, $\tilde C_g$, 
and they appear explicitly when the $Q^2 \rightarrow \infty$ limit is taken.
As a general rule, the divergent
(or soft) part of the cross section can be calculated 
from the expectation value of the quark singlet axial current between
off-shell gluon lines \cite{carlitz88,bodwin90,lech,brodsky}: 

\begin{eqnarray}
\Delta q^g (x,m_q^2,P^2,\mu^2) &=& - 2\alpha_s \mu^{-(D-4)}\int 
\frac{d^{D-2}k_\perp}{(2\pi)^{D-2}}\left[\frac{k_\perp^2 (1 - 2x) - m_q^2}
{(k_\perp^2 + m_q^2 + x(1-x)P^2)^2} \right. \nonumber \\*
&&\left. - 2\frac{D-4}{D-2}(1-x)
\frac{k_\perp^2}{(k_\perp^2 + m_q^2 + x(1-x)P^2)^2}\right],
\label{e3}
\end{eqnarray}
where $P^2 = -p^2$ is the gluon virtuality, $m_q$ is the quark mass, and 
the number of quark flavors was set to 1. For an arbitrary number of
flavors, Eq. (\ref{e3}) is multiplyed by $n_f$.
The integral can be calculated in $D$ dimensions as it stands, and the use
of the modified Minimal Subtraction ($\overline{MS}$) method to remove
the UV divergence of $\Delta q^g$  will define the
coefficient functions and parton distributions in that scheme. 
A second option is to take from the 
start the limit $D \rightarrow 4$, and use a cutoff $\mu^2$ to regularize the 
mass divergences. This is a momentum subtraction scheme, and the
anomalous gluon contribution to the first moment of $g_1^p (x, Q^2)$ will
appear\footnote{Schemes like the AB of Ref. \cite{forte96}, or JET of
Ref. \cite{leader98}, belong to this class, and we will denote this
class of schemes by $\mu$ schemes.}. 
Explicitly,

\begin{equation}
\Delta q_g^{\overline{MS}} (x,m_q^2,P^2,\mu^2) = 
\frac{\alpha_s}{2\pi}\left[(2x-1) ln\left(\frac{\mu^2}{m_q^2 + x(1-x)P^2}
\right)
- \frac{m_q^2}{m_q^2 + x(1-x)P^2} + 1 \right],
\label{e4}
\end{equation}

\begin{eqnarray}
\Delta q_g^\mu (x,m_q^2,P^2,\mu^2) &=& 
\frac{\alpha_s}{2\pi}\left[(2x-1) ln\left(\frac{\mu^2 + m_q^2 + x(1-x)P^2}
{m_q^2 + x(1-x)P^2}\right) \right. \nonumber \\*
&& \left. + (1-x) \frac{\mu^2}{\mu^2 + m_q^2 + x(1-x)P^2} 
\frac{2 m_q^2 + x(1-2x)P^2}{m_q^2 + x(1-x)P^2}\right].
\label{e5}
\end{eqnarray}
Both Eqs. (\ref{e4}) and (\ref{e5}) are dependent on the $m_q^2/P^2$ 
ratio, which is not a real problem because they are only part of the 
gluon coefficient function. What configures a problem is the attempt to 
draw conclusions about the possible anomalous gluon contribution to 
$g_1^p (x, Q^2)$ from those two equations.
The standard procedure is to look at the subtracted partonic 
cross sections:

\begin{eqnarray}
C_g^{\overline{MS}} (x, Q^2, \mu^2) &=& \tilde C_g (x, m_q^2, P^2, Q^2) 
- \Delta q_g^{\overline{MS}} (x,m_q^2,P^2,\mu^2) \nonumber \\*
C_g^\mu (x, Q^2, \mu^2) &=& \tilde C_g (x, m_q^2, P^2, Q^2) 
- \Delta q_g^\mu (x,m_q^2,P^2,\mu^2).
\label{e6}
\end{eqnarray}
The hard part of the cross sections should not depend on the $m_q^2/P^2$
ratio for $Q^2, \mu^2 >> m_q^2, P^2$, which is the reason for the neglect of 
those two variables in the left hand side of Eq. (\ref{e6}). We also 
use the label $\overline{MS}$ to denote the fact that the usual 
$\overline{MS}$ gluon coefficient function is recovered in the large
$Q^2$ limit. The same for the $C_g^\mu$ defined in a momentum 
subtraction scheme. 

In the region of low $k_\perp$, the integrands of Eq. (\ref{e3}) and of
$\tilde C_g$ are equal. Hence, the fact that the RHS of Eqs. (\ref{e6}) turns
out to be nonzero is a reflection of the large $k_\perp$ region, and of the
UV regulator of Eq. (\ref{e3}): the 
redefinition of the parton distributions, through an absortion of finite parts
of the cross section, is a matter of taste. Depending on the regulator
chosen, one can also absorb or not the axial anomaly term into the redefinition
of the parton distributions. 
Explicitly, when $Q^2, \mu^2 >> m_q^2, P^2$, we have: 

\begin{eqnarray}
C_g^{\overline{MS}} (x, Q^2, \mu^2) &=& 
\frac{\alpha_s}{2\pi}\left\{(2x-1) \left[ln \left(\frac{Q^2}{\mu^2}\right)  
+ ln \left(\frac{1-x}{x}\right) - 1 \right] + 2(1-x)\right\}, \nonumber \\*
C_g^\mu (x, Q^2, \mu^2) &=& 
\frac{\alpha_s}{2\pi} (2x-1) \left[ln \left(\frac{Q^2}{\mu^2}\right)  
+ ln \left(\frac{1-x}{x}\right) - 1 \right].
\label{e7}
\end{eqnarray}

\begin{figure}[htb]
\begin{center}
\epsfig{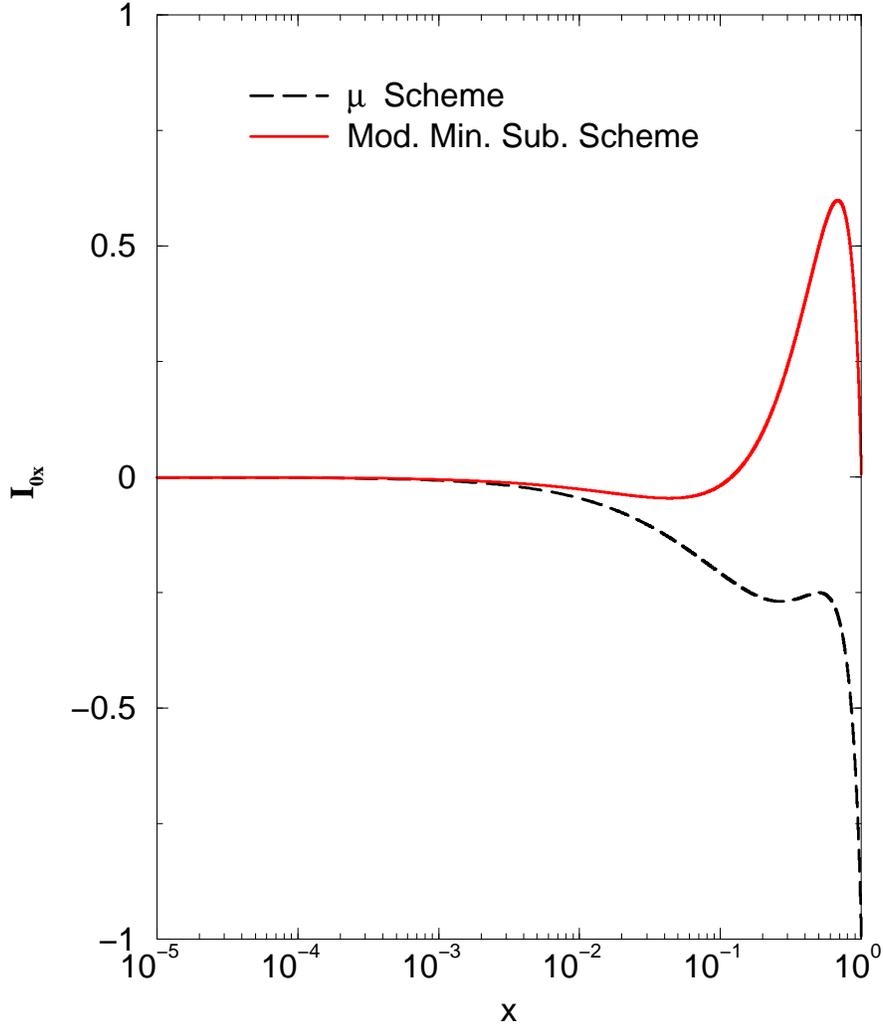}
\caption{The integral from $0$ to $x$ of the polarized gluon coefficient
function as a function of $x$, for the $\overline{MS}$ and $\mu$ schemes, 
in units of $\alpha_s/2\pi$.
schemes.}
\label{fig1}
\end{center}
\end{figure}

Contrary to repeated claims in the literature 
\cite{forte96,leader98,leader99}, the schemes discussed here have a
well defined separation of hard effects in the coefficient functions
and soft effects in the parton distribuitions. In principle, the 
polarized light quark sector is well described by both of them.
For a better understanding of both schemes, we show in Fig. \ref{fig1} 
the integrals in $x$ of Eqs. (\ref{e7}) as a function of $x$,
denoted by $I_{0x} = \int_0^x C_g^{\mu,\overline{MS}}(x,Q^2) dx$.
As is well known, 
$\int_0^1 C_g^{\overline{MS}} (x, Q^2, \mu^2) = 0$,  and 
$\int_0^1 C_g^\mu (x, Q^2, \mu^2) = -1$, in units of $\alpha_s/2\pi$.
The interesting feature is that the main contribution to both integrals
comes from the large $x$ region. In fact, $x = 0.001$ is already a good
zero, while the $x > 0.8$ region is essential to give the integrals
the value they have.

The physical structure function is indifferent to which scheme is
used to define the parton distribution and the coefficient functions.
This is expressed as:

\begin{eqnarray}
g_1^p (x, Q^2) &=& \left(\frac{1}{12}g_a^{\overline{MS}}(x) +  
\frac{1}{36}g_a^{8,\overline{MS}}(x)\right)\otimes 
C_q^{NS,\overline{MS}}(x,Q^2) \nonumber \\*
&+&
\frac{1}{9}g_a^{0,\overline{MS}}(x)\otimes C_q^{S,\overline{MS}}(x,Q^2) +
\frac{1}{9}\Delta g^{\overline{MS}}(x)\otimes 
C_g^{\overline{MS}}(x,Q^2) \nonumber \\*
&=& \left(\frac{1}{12}g_a^{\mu}(x) +  
\frac{1}{36}g_a^{8,\mu}(x)\right)\otimes C_q^{NS,\mu}(x,Q^2) \nonumber \\*
&+&
\frac{1}{9}g_a^{0,\mu}(x)\otimes C_q^{S,\mu}(x,Q^2) +
\frac{1}{9}\Delta g^{\mu}(x)\otimes C_g^{\mu}(x,Q^2).
\label{e8}
\end{eqnarray}
Although we did not write explicitly the $Q^2$ dependence of the
various distributions, we remind the reader that only the
singlet axial charge, in the $\overline{MS}$ scheme, has a 
$Q^2$ dependent first moment.

To ${\cal O}(\alpha_s)$, $C_q(x,Q^2) \equiv C_q^{NS}(x,Q^2) = C_q^S(x,Q^2) 
= \delta(x-1) + \frac{\alpha_s(Q^2)}{2\pi} C_q^{(1)}(x)$.
Using Eqs. (\ref{e7}), and the relation between the singlet axial charges
between the two schemes, $g_a^{0,\overline{MS}} (x, Q^2) = 
g_a^{0,\mu}(x,Q^2) - 
\frac{\alpha_s(Q^2)}{\pi}n_f (1-x)\otimes\Delta g (x, Q^2)$, 
the second line of Eq. (\ref{e8}) can be rewritten as:

\begin{eqnarray} 
g_1^p (x, Q^2) &=& \left(\frac{1}{12}g_a^{\mu}(x) +  
\frac{1}{36}g_a^{8,\mu}(x)\right)\otimes C_q^{\mu}(x,Q^2) \nonumber \\*
&+&
\frac{1}{9}g_a^{0,\overline{MS}}(x)\otimes C_q^{\mu}(x,Q^2) +
\frac{1}{9}\Delta g^{\mu}(x)\otimes C_g^{\overline{MS}}(x,Q^2),
\label{e9}
\end{eqnarray}
where the term $\alpha_s^2/2\pi^2 (1-x) \otimes\Delta g(x) 
\otimes C_q^{(1)}(x)$ was disregarded.
We can now relate the remaining distributions and coefficient functions
in the two schemes in the following way:

\begin{eqnarray}
C_q^\mu (x) &=& C_q^{\overline{MS}} (x) + \delta C_q(x) \nonumber \\*
\Delta g^\mu (x) &=& \Delta g^{\overline{MS}} (x) + \delta g(x) \nonumber \\*
\frac{1}{12}g_a^\mu (x) +  \frac{1}{36}g_a^{8,\mu} (x) &=& 
\frac{1}{12}g_a^{\overline{MS}}(x) +  \frac{1}{36}g_a^{8,\overline{MS}}(x)
+ \delta q (x),
\label{e10}
\end{eqnarray}
where $\delta C_q (x)$, $\delta g (x)$ and $\delta q (x)$ are some general 
functions, of ${\cal O}(\alpha_s)$. Their specific form is not of interest 
to us at this given moment. However, the use of Eqs. (\ref{e10}) in 
Eq. (\ref{e9}), and the requirement that Eq. (\ref{e8}) is satisfied, 
produces the following consistency relations:

\begin{eqnarray}
&& \left(\frac{1}{12}g_a^{\overline{MS}}(x) 
+ \frac{1}{36}g_a^{8,\overline{MS}}(x)
+\frac{1}{9}g_a^{0,\overline{MS}}(x) + \delta q(x) \right)
\otimes \delta C_q (x) \nonumber \\*
&& + \delta q (x)\otimes C_q^{\overline{MS}}(x) + 
\delta g (x) \otimes C_g^{\overline{MS}}(x) = 0,
\label{e11}
\end{eqnarray}
\begin{eqnarray}
&& \left(\frac{1}{12}g_a^{\mu}(x) +  \frac{1}{36}g_a^{8,\mu}(x)
+ \frac{1}{9}g_a^{0,\mu}(x) - \delta q(x) \right)
\otimes \delta C_q (x) \nonumber \\* 
&& + \delta q (x)\otimes C_q^{\mu}(x) + 
\delta g (x) \otimes 
\left(C_g^{\mu}(x) - \frac{\alpha_s}{\pi} n_f (1-x)\right) = 0.
\label{e12}
\end{eqnarray}
The first moment of Eq. (\ref{e11}) 
certainly respects the equality, as 
$\int_0^1 C_g^{\overline{MS}}(x) dx = 0$ and
$\int_0^1 \delta q(x) dx = \int_0^1 \delta C_q (x) dx = 0$ because of the 
conservation of the nonsinglet axial current
\footnote{As the change of the 
coefficient function is dictated by the change of scheme of the anomalous
dimension, and the first moment of the nonsinglet anomalous dimension is
zero due to current conservation, it follows that 
$\int_0^1 \delta C_q (x) dx = 0$.}. 
The conservation of the nonsinglet axial current also imposes, from the 
first moment of Eq. (\ref{e12}), that $\int_0^1 \delta g(x) dx = 0$, because 
$\int_0^1 C_g^{\mu} (x) dx = -n_f \alpha_s/2\pi \Delta g$. It follows that 
the first moment
of the polarized gluon distribution is the same in the $\overline{MS}$ and
$\mu$ schemes, independent of whether $\Delta g$ contributes or not to the
first moment of $g_1^p (x,Q^2)$, up to the $(\alpha_s/2\pi)^2 \Delta g$ 
corrections we neglected before. This result is consistent
with the fact that $\Delta g$ starts contributing to $g_1^p (x, Q^2)$ at
${\cal O}(\alpha_s)$ only.

\begin{figure}[htb]
\begin{center}
\epsfig{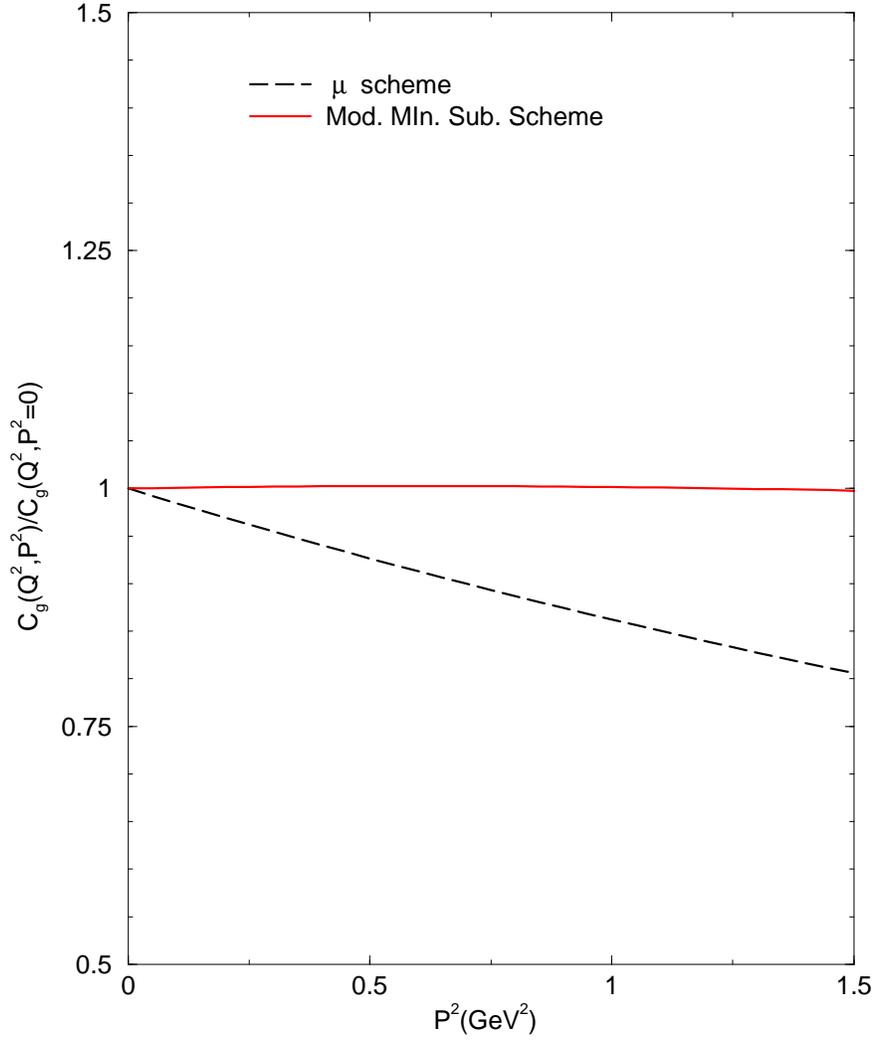}
\caption{The integrated (in $x$) polarized gluon coefficient function
in the $\overline{MS}$ and $\mu$ schemes, as a function of $P^2$.}
\label{fig2}
\end{center}
\end{figure}

Although both schemes are, in principle, equally good to describe 
$g_1^p (x, Q^2)$ in the light quark sector, we should also look at
their behavior when heavy quarks are introduced. In particular, we do not
want the hard part of the cross sections to depend on $P^2$ once
the mass of the heavy quark and the factorization scale are 
fixed.
To investigate that, we calculate Eqs. (\ref{e6}) as a function of 
$P^2$ for the charm quark, with $m_c = 1.5 \; GeV$ and $Q^2 = \mu^2 = 
10 \; GeV^2$.
The resulting curves are shown in Fig. \ref{fig2}, normalized by the
coefficient functions calculated with $P^2 = 0$.
It is clear that $C_g^{\overline{MS}}$ is independent of $P^2$ in the
range $0 \leq P^2 \leq m_c^2$. The same is not true for $C_g^{\mu}$, 
which shows a strong dependence on $P^2$. 

Numerically, 
$\int_0^1 C_g^{\overline{MS}} (x, m_q^2 = 2.25 \; GeV^2, Q^2 = 10 \; GeV^2) dx
\approx 0.4$, in units of $\alpha_s (Q^2)/2\pi$.
Of course, this integral changes with 
$Q^2$, going to zero as $Q^2 \rightarrow \infty$, but it is 
independent of $P^2$ for fixed $Q^2$. 
Hence there is a well defined contribution from gluons 
to the first moment of $g_1^p (x, Q^2)$, in the $\overline{MS}$ scheme,
which appears because of the relatively large mass of the charm quark.
On the other hand,
$\int_0^1 C_g^\mu (x, m_q^2 = 2.25 \; GeV^2, Q^2 = 10 \; GeV^2)dx$
ranges from $\sim  - 0.18$, at $P^2=0$,  to $\sim -0.135$,
at $P^2 = m_c^2$. Although the difference is not numerically significant 
($(0.18 - 0.135)(\alpha_s/2\pi)\Delta g \sim  0.001 \Delta g$) 
as long as $\Delta g$ is not very large,
the use of the $\mu$ schemes is, to some degree, damaged. 

The inclusion of heavy quarks in the framework of perturbative calculation of 
structure functions have received great attention in the recent literature
\cite{acot,me98,kretzer,buza,thorne,collins}. These works have focused in the 
development of shemes that interpolate the pure photon-gluon fussion
calculation from the region where $Q^2 \sim m_h^2$, to the usual massless
approach (when $Q^2 >> m_h^2$). A fundamental point of these schemes is
that the heavy quark is treated as a massless parton in the Altarelli-Parisi 
evolution equations, which will have $n_f + 1$ active flavours, while the
quark mass dependence is fully kept in the graphs containing 
the heavy quark lines in the calculation of the coefficient functions.
These schemes are generally refered to as Interpolating 
Schemes (IS).

The coefficient functions in Eq. (\ref{e6}) incorporate the full mass
corrections, and are reduced to the massless case in the limit of
large $Q^2$. Hence, they are suitable for the calculation of the
polarized structure functions for $Q^2 \sim m_h^2$ and $Q^2 >> m_h^2$, 
in the spirit of the IS.
As in the IS the light and the newly introducded heavy quark
distributions should be defined in the same scheme, and as the calculation 
of the $C_g^\mu$ for a heavy quark is ambiguous, it follows that,
strictly, the $\overline{MS}$ scheme is formally superior to the 
$\mu$ scheme for the calculation of $g_1^p (x, Q^2)$.

As a last remark, we want to stress that the amount of polarized 
heavy quark in the proton is not given by the integral in $x$ of 
Eqs. (\ref{e4}) and (\ref{e5}), or from the integral of $\tilde C_g (x,Q^2)$
for a given quark mass.
From them, one would conclude that
$\int_0^1 \Delta q_g^{\overline{MS}} (x,m_q^2,P^2,\mu^2)dx = 0$ for 
$\mu^2 = m_q^2 >> P^2$, while 
$\int_0^1 \Delta q_g^{\mu} (x,m_q^2,P^2,\mu^2)dx = \alpha_s/4\pi$
in that same limit.
In a framework where heavy quark mass effects are systematically 
included, one should introduce\footnote{If the heavy quark contribution to
$g_1^p(x,Q^2)$ is calculated through the photon-gluon fussion proccess only,
and its higher order corrections, a polarized heavy quark distribution
is never introduced.} 
a polarized heavy quark distribution 
in the proton ($\Delta h (x,Q^2)$) at the factorization 
scale $\mu^2$, with $\Delta h(x,\mu^2)=0$\footnote{It is assumed that   
there is no intrinsic heavy quark polarization in the proton. See
Ref. \cite{bass99} for a different point of view.}.
As we saw here, both $\overline{MS}$ and $\mu$ schemes are suitable 
for this purpose once Eqs. (\ref{e6}) are given, although, in principle,
the $\overline{MS}$ scheme has the technical advantage of having a
$P^2$ independent gluon coefficient function in the heavy quark sector.
\newline
I would like to thank X. Ji and A. W. Thomas for valuable discussions.
This work was supported by FAPESP (under contracts 96/7756-6 and 98/2249-4).

\addcontentsline{toc}{chapter}{\protect\numberline{}{References}}

\end{document}